\documentclass[modern]{aastex63}
\usepackage{amssymb, amsmath, framed}
\usepackage{bm}
\expandafter\ifx\csname package@font\endcsname\relax\else
 \expandafter\expandafter
 \expandafter\usepackage
 \expandafter\expandafter
 \expandafter{\csname package@font\endcsname}
\fi
\usepackage{graphicx}	
\usepackage{amsmath}	
\usepackage{amssymb}	

\usepackage[T1]{fontenc}
\usepackage[utf8]{inputenc}

\begin{document}

\title{Longevity is the key factor in the search for technosignatures}

\correspondingauthor{Amedeo Balbi}
\email{balbi@roma2.infn.it}

\author[0000-0002-3929-6932]{Amedeo Balbi}
\affiliation{Dipartimento di Fisica, Universit\`a degli Studi di Roma ``Tor Vergata'', Via della Ricerca Scientifica, 00133 Roma, Italy}
\author[0000-0002-6634-1321]{Milan M. \'Cirkovi\'c}
\altaffiliation{Also at Future of Humanity Institute, Faculty of Philosophy, University of Oxford, Suite 8, Littlegate House, 16/17 St Ebbe's Street, Oxford, OX1 1PT, UK}
\affiliation{Astronomical Observatory of Belgrade, Volgina 7 11000 Belgrade, Serbia}

\begin{abstract}
It is well-known that the chances of success of SETI depend on the longevity of technological civilizations or, more broadly, on the duration of the signs of their existence, or technosignatures. Here, we re-examine this general tenet in more detail, and we show that  its broader implications were not given the proper significance. In particular, an often overlooked aspect is that the duration of a technosignature is in principle almost entirely separable from the age of the civilization that produces it. We propose a classification scheme of technosignatures based on their duration and, using Monte Carlo simulations, we show that, given an initial generic distribution of Galactic technosignatures, only the ones with the longest duration are likely to be detected. This tells us, among other things, that looking for a large number of short-lived technosignatures is a weaker observational strategy than focusing the search on a few long-lived ones. It also suggests to abandon any anthropocentric bias in approaching the question of extraterrestrial intelligence. We finally give some ideas of possible pathways that can lead to the establishment of long-lived technosignatures. 
\end{abstract}


\section{Introduction}
In recent years, the discovery of a large number of potentially habitable exoplanets, as well as several other breakthroughs in astrobiology, has motivated a resurgence of interest in the search for extraterrestrial intelligence (SETI). The scope of the investigation has surpassed the boundary of traditional searches for radio signals, expanding to encompass the more general category of `technosignatures'---i.e.\ of remotely detectable signs that can be linked to the activity of technological civilizations on other planets \citep{Tarter2006,NASA2018,Wright2019}.

Ever since the early days of SETI, it was clear that the longevity of technological civilizations was crucial in evaluating the chances of finding evidences of their existence. In fact, longevity was a key factor in the Drake equation estimating the present number of communicating civilizations (\citet{Drake1965}; see also \citet{Drake1992,Dick1996}). Strangely enough, this point has often been a target for vehement criticism by the opponents of SETI who, profusely helped by the media, saw the project in a quasi-religious light. Painting of SETI as a ``search for alien gods'' by \citet{Tipler1981}, \citet{Mayr1993}, \citet{Basalla2006} or \citet{Kukla2010}, has already inflicted much damage on both the image and funding of the whole enterprise. Future historians of science will perhaps need to elaborate on why the rather commonsense intuition that, since we have just emerged on the cosmic scene, it is probable that any potential targets will be older than us, caused so much outrage.  

In this paper, we wish to re-examine the role of longevity of potential SETI targets for several reasons which are tightly connected with the overall reorientation of the field, which include enlarging the scope of the search to the more general class of technosignatures. In particular, the strategies which have been conceived in the early days of SETI in 1960s and 1970s for the search of Earth-like inhabited planets emitting intentional radio-signals need to be embedded in a wider spectrum of approaches and strategies. The increased interest in SETI studies characterizing the last decade has already led to generalizing efforts in both the theoretical and observational domain \citep[e.g.,][]{Wright2014a, Wright2014b, Zackrisson2015, Gillon2014}.

The outline of the paper is as follows. First, in Section~\ref{1}, we will briefly review the role of longevity in classical SETI studies and make the very notion more precise. Next, in Section~\ref{2}, we will propose a more general approach that can overcome some of the shortcomings of the traditional treatment, and can be applied to a more general set of technosignatures. In Section~\ref{3}, we will use this conceptual framework to explore the longevity features of the sub-set of technosignatures that can be detected (if it is not empty), while in Section~\ref{4} we will propose a classification scheme of technosignatures based on their longevity.  Finally, in Section~\ref{5}, we will introduce some illustrative models of longevity and show how they can affect the chances of success of technosignatures searches. 

\section{The classical treatment}\label{1}
The classical approach to estimating the present number of technological civilizations in a volume around Earth (usually, the Galaxy) is based on the famous Drake equation \citep{Drake1965} which, in its most basic form, can be written as:
\begin{equation}
\label{Drake}
N=\Gamma  L,
\end{equation} 
where $\Gamma$ is the average rate of appearance of communicating civilizations and $L$ is their average longevity. In steady-state, $N$ represents the average number of communicating civilizations at any epoch in the volume under examination. Usually, the rate of appearance $\Gamma$ is written as the product of the probabilities attached to the various processes that are deemed necessary for the arising of communicating civilizations. Here, however, we are not interested in a discussion of such factors, which is the subject of countless previous works resulting in disparate estimates of $N$. Rather, we wish to bring the attention to the fact that the equation is, in essence, the ratio of two temporal scales, i.e.\ $\Gamma^{-1}$ and $L$. These two quantities are independent of each other: however, their relative amplitude determines the value of $N$. 

There has always been some ambiguity regarding the exact meaning of $L$. Usually, it has been defined operationally as the average duration of the communication phase of an extraterrestrial civilization \citep[e.g.,][]{Drake1992}. It has occasionally been suggested on that basis that for humanity $L_\oplus \approx 100$ years, since the development of radio by Tesla and Marconi. However, this raises at least three types of problems: (i) how should the averaging be performed? (ii) what kind of communication? (iii) what about extinct civilizations? The usage of radio as the normative manner of communication made sense in the early days of SETI, but is hardly supported nowadays when even human civilization relies progressively less on radio communication than it was the case a few decades back. While there might still be some advantages in interstellar radio \citep{Hippke2018}, today we clearly see that diverse types of technosignatures require different effective definitions of the ``communication window'', hence different effective values of $L$.

While it is of necessity partially anthropocentric, there is the obvious possibility that many extraterrestrial civilizations will be extinct (like the majority of terrestrial human civilizations) and yet be in principle detectable through their technosignatures. While the complex topic of ``interstellar archaeology'' is mostly beyond the scope of the present study \citep[e.g.,][]{Carrigan2012, Davies2012}, the distinction between the longevity of isolated technosignatures and the longevity of either existent or extinct parent civilization should be kept in mind. At least one recent study \citep{Stevens2016} investigated the possibility of detecting technosignatures associated with the extinction of an extraterrestrial civilization. 

The most serious theoretical confusion about the relation of $N$ and $L$, however, is related to the averaging procedure. Even neglecting the slow processes driven by cosmological and astrophysical evolution (e.g., changes in the terrestrial planet formation rates), there could be all kinds of spatial and other parameter variations which has not been seriously modeled thus far. There is also a puzzling, partly philosophical, question: how much could $L$ be extended into the future from a fiducial moment in time (e.g., the present)? Any specific numerical value of $L$ is by necessity an average. However, the same average value can be produced by drastically different astrobiological situations; for example, depending on whether most civilizations over which the averaging is performed are located in our past or in our future. This is a consequence of rejecting temporal Copernicanism \citep{Cirkovic2020}. Similar to the value of money (the value of \$100 is very different across the span from a homeless guy to a Silicon Valley billionaire!), the contribution to detectability of changing longevity by some $\Delta L$ is itself a function of $L$ and also likely to change with cosmic time. Hence the need for simplifying assumptions. 

An important point to note is that in classical SETI $N$ was interpreted as the number of technological civilizations that are actively communicating, at any time, over interstellar distances (usually, through radio waves). However, it can be easily generalized to indicate any sign of technological activity that can be detected by means of remote investigation. Thus, from now on we will refer to $N$ as the number of {\em detectable} technosignatures. We should point out from the outset that we are only interested in what can be detected {\em in principle}, in a sense that will be clearer later. Thus, we will ignore aspects such as signal strength, observational limitations, technical capabilities, and so on. In other words, for our purposes the possibility of detecting a technosignature is a binary variable: either we can or we cannot.  We shall also ignore those necessarily existent borderline cases in which the artificial hypothesis for the detected signal is the best explanation, but not overwhelmingly so. While the latter cases might be the most probable outcome of a future comprehensive SETI effort, it is reasonable to assume that follow-up observations will be able to set the record straight one way or another; hence our approach does not lose generality, up to a time-stamp.

If the rate of appearance $\Gamma$ is constant in time, its value can be estimated as the total number of technosignatures that appeared over the history of the Galaxy $N_{\rm tot}$, divided by the age of the Galaxy $T_G$. Thus, the number of presently detectable technosignatures is simply:
\begin{equation}
N = N_{\rm tot} \frac{L}{T_G}. \label{longdrake}
\end{equation}
It is apparent that the number of technosignatures that we can detect is a fraction of the total number that ever existed: the fraction is precisely $L/T_G$. Because $T_G\sim 10^{10}$ years, $L/T_G$ is presumed to be generally small; any specific precondition imposed on the origination of technosignatures, like the necessity of terrestrial planets for biological evolution, will act to reduce the fraction. This is the quantitative argument that justifies one of the most widely cited assertions of classical SETI, i.e.\ that the chances of finding ETIs depend on the average longevity of technological civilizations. (In fact, it is well-known that Frank Drake himself used to equate $N$ to $L$.)  

However, Eq.~(\ref{longdrake}) can be also interpreted in a different manner: unless $L$ is large (that is, comparable to $T_G$) a very large $N_{\rm tot}$ is needed in order for $N$ to be greater than unity. That is, there must have been a large number of technosignatures over the history of the galaxy, to have just a few detectable ones at present. 

\begin{figure*}[ht]
\includegraphics[width=\textwidth]{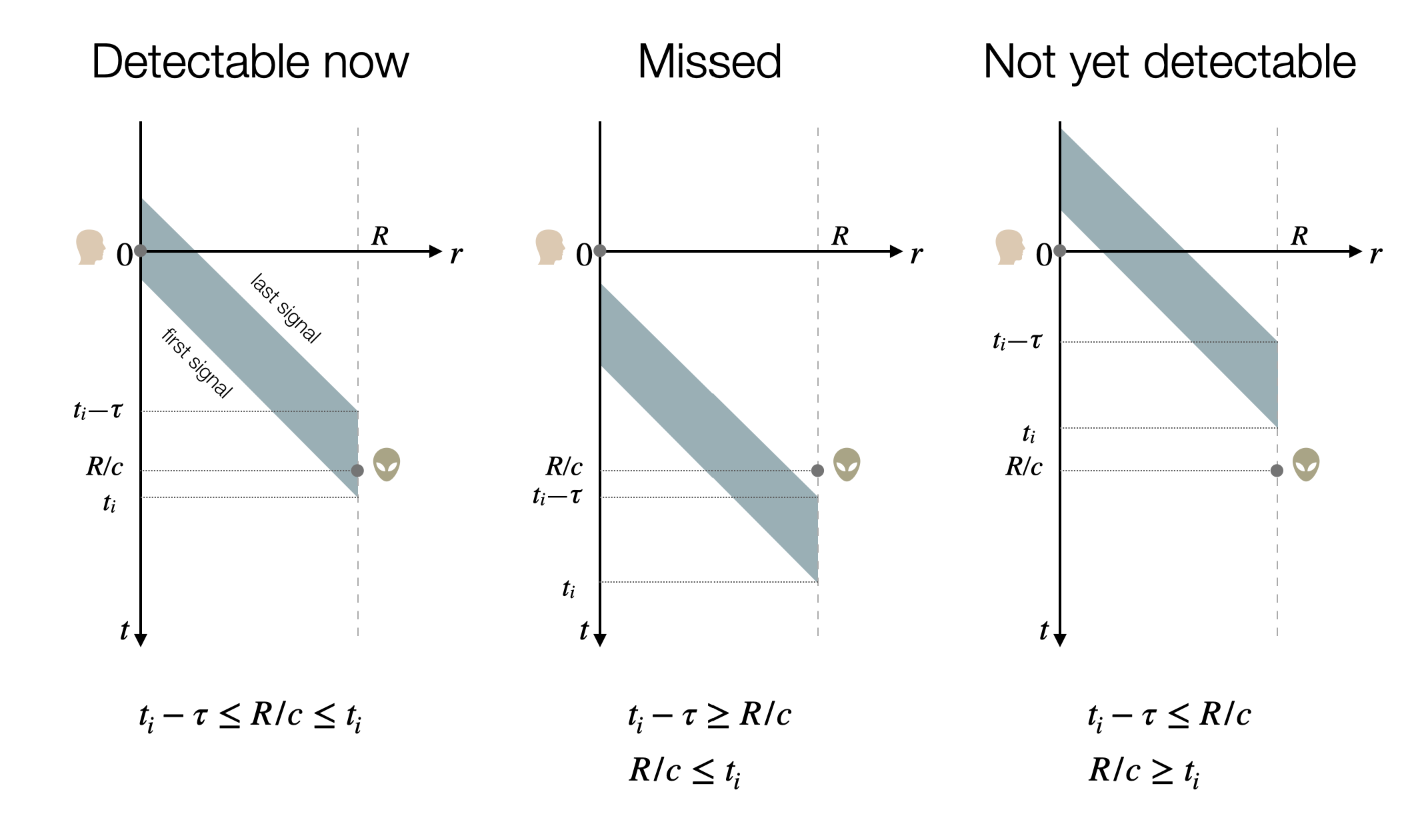}
		\caption{Causal structure of technosignature detection. The Earth is located at $r=0$, and the present time is $t=0$ (with the temporal coordinate taken as positive and growing in the past). A technosignature at distance R is detectable today only if its time of appearance $t_i$ and duration $\tau$ satisfy the constraint of Eq.~\ref{causal} (left panel). The technosignature is not detectable either if its last evidence reached us in the past (center panel) or if it has not yet reached us (right panel)}\label{contact}
\end{figure*}

\section{A more general approach}\label{2}

Despite its doubtless utility and widespread application, the Drake equation has a number of shortcomings. For the present discussion, the most relevant one is the lack of an explicit temporal dependence \citep{Cirkovic2004}. In other words, both $\Gamma$ and $L$ are average quantities, and there is an implicit assumption that $N$ is stationary over the history of the Galaxy.  There are good reasons to believe that this is not the case. Of course, it is unrealistic to assume that $\Gamma$ is constant with cosmic time. Even if we limit ourselves to the last $\simeq 10$ Gyr of existence of thin disk Pop I stars which are likely to harbour the predominant fraction of all possible habitats for intelligent species, their rate of emergence is likely to be very nonuniform. One obvious source of nonuniformity is the changing rate of emergence of planetary habitats, as first established by \citet{Lineweaver2001} and subsequently elaborated by \citet{Behroozi2015}, as well as by \citet{Zackrisson2016}. This nonuniformity can be precisely quantified today and some contemporary astrobiological numerical simulations have taken it into account \citep{Dosovic2019}. 

All in all, it is reasonable to presume that the habitability of the universe is a function of time \citep{Loeb2016, Dayal2016}. Thus, it would make sense to consider the rate of appearance of technosignatures as varying in time, $\Gamma=\Gamma(t)$. Similarly, averaging longevity over time might cause important details to go missing. The history of humanity, short and simplistic as it is, gives us reason to be quite skeptical, at the very least, with regard to the averaging procedure. For example, as noted in \citet{Cirkovic2019}, older artifacts like the Pyramids may last longer than the newer ones, so the averaging operation would be problematic. Several difficulties are likely to beset any attempt to judge the longevity of a civilization by the age/rate of appearance of their technosignatures. While the Pyramids of Egypt were indeed produced by a relatively long-lived civilization, there is no clear correlation here, since some of other durable artifacts of human past were produced by relatively short-lived cultures (e.g., the walls of Cuzco). The rate of artifact destruction on Earth has, in general, been a function of both geographical location and time, even if we limit ourselves to non-human causative agents; selection bias exists, for instance, against artifacts in floodplains, in tectonically active areas or on coastlines submerged since the end of the last ice age. The analog rates of attrition or loss for cosmic technosignatures have not been considered so far.

Other nonuniformities are likely to exist, though, and are more difficult to quantify. Historically, these have been associated with the contingency and opportunism of evolution connecting the existence of habitats with the evolution of intelligence and civilization. This important point has been emphasized by critics of SETI since the onset of the first searches \citep{Simpson1964}, but unfortunately has not been sufficiently addressed by the SETI community. Evolution tends to invent new ways of solving persistent problems of populations in given ecosystems, which often leads to abandoning old procedures and toolkits (some could subsequently be exapted or coopted for other purposes; cf. \citet{Gould1982}). An analog of this within the framework of cultural evolution could correspond to the loss of incentives and motivations for producing a particular class of artifacts. While modern-day humans could, arguably, produce even larger pyramids than those of the III-VI dynasties of ancient Egypt, they clearly do not engage in such projects, since there is no corresponding motivation. (Note, however, {\em cultural exaptation} of the surviving ones as tourist attractions.) We shall return to it in the discussion below in connection with the long-living (Type C) technosignatures. 

To properly address the temporal dependences implicit in the estimate of $N$, a more flexible approach can be adopted, as proposed in \citet{Balbi2018a} \citep[see also][]{Grimaldi2017,Lares2020, Grimaldi2020}. This relies on noticing that any technosignature that we can observe must be located within our past light cone; that is, we can detect a technosignature at distance $R$ from Earth, at present, if and only if:
\begin{equation}\label{causal}
t_i-\tau \le R/c\le t_i
\end{equation}
where $t_i$ is the initial time of the technosignature (taken as positive and growing towards the past, starting from the present epoch $t_0=0$) and $\tau$ is its overall duration, that is, the time during which it remains detectable via causal signals (see Fig.~\ref{contact}).

There is an obvious connection between this formalism and the one of the Drake equation: given an ensemble of technosignatures, the distribution of $t_i$ will be related to the rate of appearance $\Gamma$ (e.g., if $\Gamma$ is constant, $t_i$ will be uniformly distributed in $[0,T_G]$), while $L=\langle \tau \rangle$, where the bracket represent the average over time. However, Eq.~(\ref{causal}) is more general and it is better suited to treat cases where there is a temporal variation in $\Gamma$ or $\tau$ or both. In particular, the number of technosignatures that can be detected at present, $N$, can be estimated by simulating  $N_{\rm tot}$ random values of $t_i$, $\tau$ and $R$, drawn from appropriate probability distributions, and then counting only those that meet the causal requirement of Eq.~(\ref{causal}). In other words, Eq.~(\ref{causal}) acts a filter that selects, among all  technosignatures that appeared during cosmic history, only the ones that we can in principle observe. We should point out that this applies regardless of the number and distribution of technosignatures existing in the universe: {\em every single tecnosignature that we can detect has to pass the filter}.

This does not preclude the existence of other filters, nor of kinds of technosignatures with characteristic timescales set by velocities other than $c$. (Examples in this sense would include transfer of information by transiting occulters, artificial sources of particles which propagate slower than light, interstellar probes, and so on.) However, as long as special relativity holds, all these other filters are contingent on the application of the one in Eq.~(\ref{causal}).

The existence of this filter suggests to look at the search for technosignatures from a different perspective than the one usually adopted. While the standard goal of most theoretical studies  on extraterrestrial intelligence was to guess the number $N$, we might shift the focus from quantity to {\em quality} and ask instead: what is the most likely {\em type} of technosignature that we can detect?

\section{What type of technosignatures can we detect?}\label{3}

As already pointed out in \citet{Balbi2018a, Balbi2018b}, the causal constraint of Eq.~(\ref{causal}) introduces a strong correlation between two otherwise independent timescales, namely $t_i$ and $\tau$. This is due to the fact that the time $R/c$ is $\lesssim 10^5$ years for any location $R$ with the Milky Way. Then, although in principle both $t_i$ and $\tau$ can assume values between 0 and $T_G$, the difference $t_i-\tau$ has to be small (i.e.\ at most $\sim 10^5$ years) for any technosignature we can detect. Another way to state this is that, because the crossing time of the Galaxy is $\sim 10^5$ years, any technosignature that ceased to be `active' before $\sim 10^5$ years ago is not observable at present from our location. 

\begin{figure*}[ht]
\includegraphics[width=\textwidth]{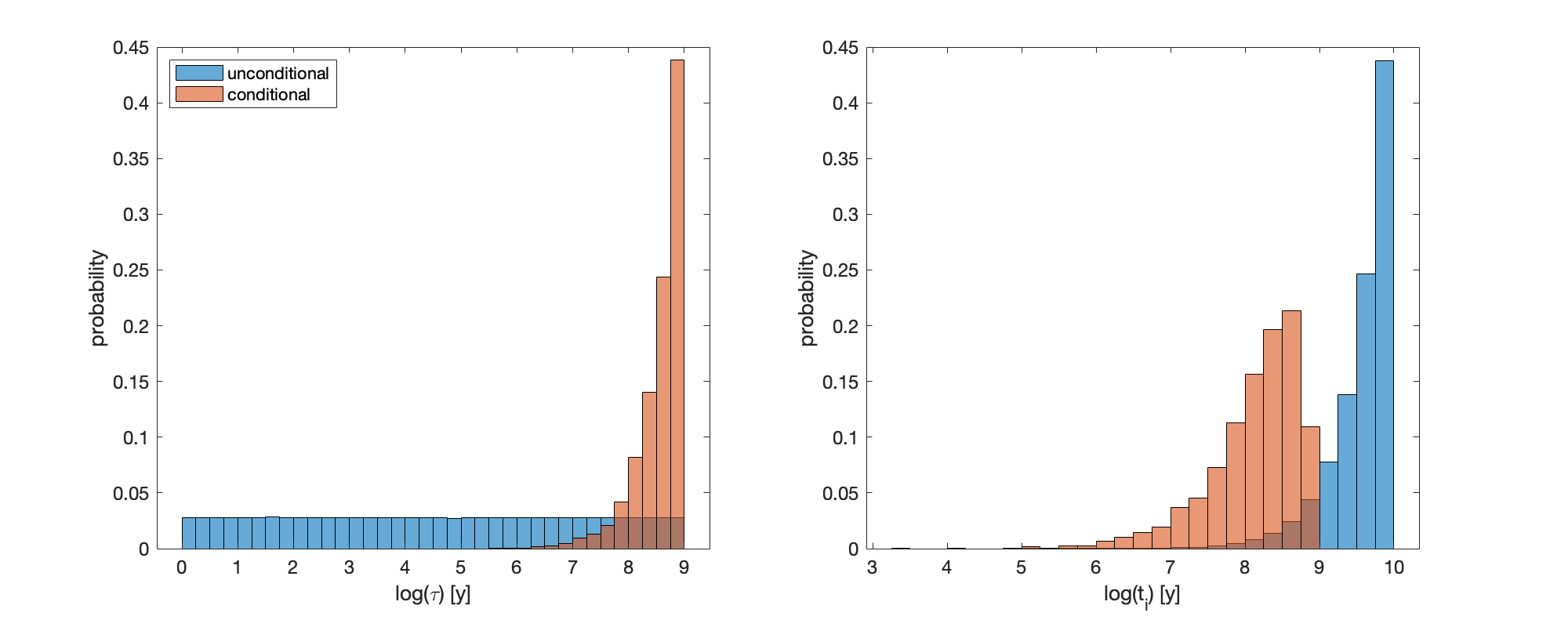}
\includegraphics[width=\textwidth]{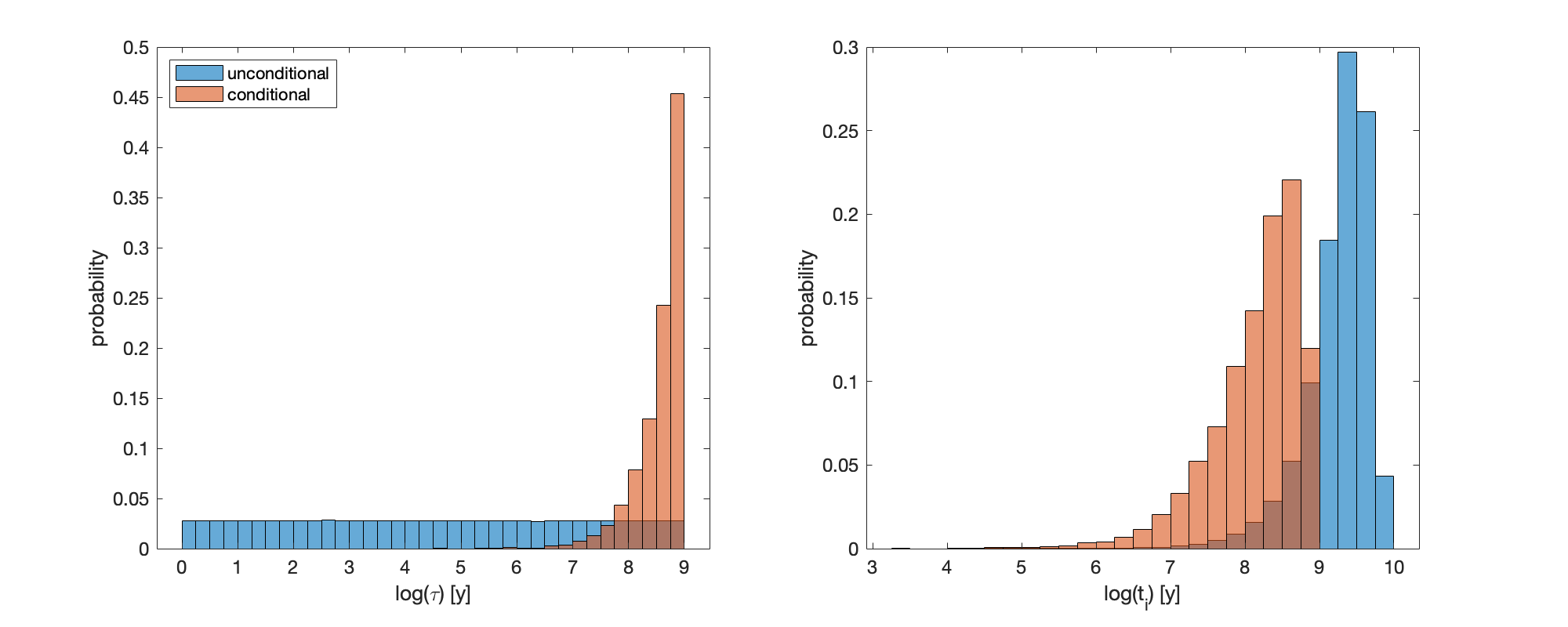}
		\caption{Each row shows the unconditional probability distribution of longevities $\tau$ (left) and time of appearance $t_i$ (right) and the probability distribution of the same quantities conditioned to detection. In both rows, the unconditional probability distribution for $\tau$ is a log-uniform, while $t_i$ is uniformly distributed in the upper row (Model 1 in the text) and normally distributed in the lower row (Model 2 in the text).}\label{conditional}
\end{figure*}

This has interesting consequences, because it can give us some insight on the following question: if we detect a technosignature at distance $R$, what should we expect its longevity $\tau$ to be?\footnote{While this manuscript was in preparation, this issue was addressed in a Bayesian framework by \citet{Kipping2020}, although no explicit use of the causal constraint required for detection was made.} A priori, we have no way of telling what the maximum duration of an intelligent civilization (or, more generally, of its technological artifacts) should be, so $\tau$ is essentially unconstrained. However, once we condition to the fact that we {\em detected} the technosignature, we select some values of $\tau$ as more probable. The minimum longevity for a detectable technosignature descends directly from Eq.~(\ref{causal}):
\begin{equation}
\tau\ge t_i-R/c,\label{longevity}
\end{equation}
so that if we treat both $\tau$ and $t_i$ as random variables, the most likely value of $\tau$ for a detectable technosignature will be linked to the distribution of $t_i$. 

We investigated this issue quantitatively, by simulating a sample of $10^6$ random technosignatures, uniformly distributed in a sphere of radius $R_E=10^3$ light years  surrounding Earth. We assumed that $t_i$ and $\tau$ are drawn from probability distributions $P_{t_i}(t_i)$ and $P_\tau(\tau)$. The sub-sample of detectable technosignatures will then have conditional probability distributions $p_{t_i}(t_i)=P_{t_i}(t_i\vert D)$ and $p_\tau(\tau)=P_{\tau}(\tau\vert D)$, where $D$ indicates the causal constraint of Eq.~(\ref{causal}).

To explore the dependence on the distribution of $t_i$, we made two different choices for the unconditional probability $P_{t_i}(t_i)$. The first (Model 1) is a uniform distribution in the interval $[0,T_G]$: this would correspond to a constant appearance rate $\Gamma$, as in the standard Drake equation. The second (Model 2) is a more realistic distribution derived from the assumption that the ages of planets suitable to host complex life follow a normal distribution with mean $5.5$~Gy and standard deviation 2~Gy \citep{Lineweaver2004}. We then used as a fiducial interval for technosignatures to appear on a suitable planet the time it took for the human species to evolve, $t_{\rm evol} =4\pm 1$ Gy. Having no idea on the distribution of $\tau$, we took as the unconditional probability $P_\tau(\tau)$ a log-uniform function, i.e.\ we assigned the same probability to any order of magnitude of $\tau$ in the interval $[0,10^9]$ years. After producing random samples of technosignatures according to the aforementioned probability distributions, we select only the technosignatures that conform to the prescription of Eq.~(\ref{causal}), and are therefore detectable today: we then use their values of $t_i$ and $\tau$ to build the conditional probability distributions for such parameters. 

Figure~\ref{conditional} shows the conditional probability for $\tau$ and $t_i$ derived from our simulations. Two results are evident. The first is that, starting from a sample of technosignature with no preference on the value $\tau$, the subset of technosignatures that can be detected from Earth have very large longevities. In other words, the biased distribution emerges from an unbiased distribution in the course of the Bayesian process \citep[cf.][]{Cirkovic2010}. The second conclusion is that this fact is robust with respect to different assumptions on the distribution of $t_i$. As a further check of this result, we also simulated a situation where the appearance of our civilization on Earth is entirely typical (Model 3), so that $t_i$ is normally distributed with mean corresponding to the present time $t_0=0$. This left the conditional probability for $\tau$ basically unchanged. We also checked that enlarging the radius of the observed volume to $R_E=10^5$ light years, i.e.\ encompassing the entire Galaxy, has very little effect. Finally, we changed the unconditional distribution of $\tau$ from a log-uniform to a linear-uniform (giving equal probability to any value in the interval $[0,10^9]$ years), finding no noticeable difference in the final result.  A strong general conclusion of our simulations is that the minimum value of $\tau$ compatible with detection is of order $\sim 10^6$ years. 

\begin{figure}[ht]
\center{\includegraphics[width=\columnwidth]{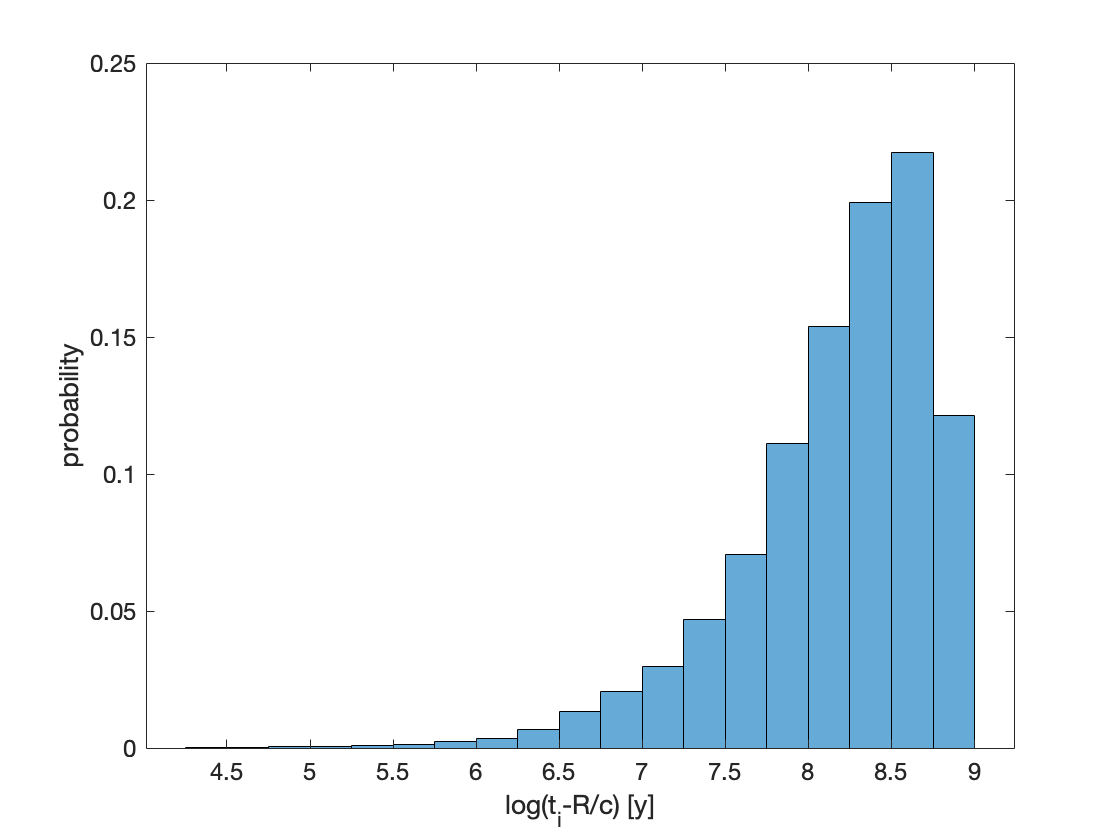}}
		\caption{The distribution of the age of technosignatures that passed the detectability constraint of Eq.~\ref{causal}, for a initial random set with log-uniform distribution in $\tau$ and normal distribution in $t_i$ (Model 2 in the text).}\label{age}
\end{figure}

We should point out that, in our formalism, $\tau$ is the projected longevity of a technosignature at the time when it first appears. In other words, the termination time of the technosignature, $t_i-\tau$, can be in the future (that is, negative times). Instead, the total time that a given technosignature has been active before detection, i.e.\ its current `age', is given by $t_a\equiv t_i-R/c$. This is shown in Figure~\ref{age}, for the subset of detectable technosignatures of our Model 2. Although the youngest technosignatures in the sample are only a few tens of thousand years old, these are just outliers: the vast majority of detectable technosignatures is very old, with a median age exceeding $10^8$ years.

Our approach could be generalized rather easily to extragalactic SETI, which has gained currency in recent years with several innovative observational searches \citep{Wright2014a, Wright2014b, Zackrisson2015}. In that case, we would need to take into account cosmological models and especially the cosmological look-back time, that is the difference between the age of the Universe now ($\approx 13.77$ Gyr) and the age of the Universe when a signal was emitted from an object at redshift $z$. In the general case this is given by:
\begin{equation}
t(z) = \frac{1}{H_0} \int_0^{z} \frac{dz}{(1+z) E(z)},
\end{equation}
where $H_0$ is the Hubble constant and the function $E(z)$ is defined as:
\begin{equation}
E(z) = \sqrt{\Omega_m (1+z)^3 + \Omega_k (1+z)^2 + \Omega_\lambda}.
\end{equation}
Here, $\Omega_m$, $\Omega_k$, and $\Omega_\lambda$ are the density fractions of matter, curvature, and dark energy, respectively. Eq.~\ref{longevity} above would need to be modified to substitute $R/c$ with the look-back time. This will suppress technosignatures in more distant galaxies, since we see them at too early stages of their astrobiological evolution. 

So far, we have discussed what is the most likely type of tecnosignatures that {\em we can detect} (in the unequivocal sense discussed above): but this does not tell us anything about what is the most likely type of technosignatures that {\em exists}. It may be possible that, in reality, the set of technological civilizations or artifacts with $\tau>10^6$ years is empty. Alternatively, while very old technosignatures could indeed exist, their unequivocal detection might be impossible due to the signal-to-noise ratio declining sharply with time (for example, due to the interstellar scintillation for radio or galactic magnetic field distortions for cosmic-ray particle signals). Eventually, the success of any search for evidences of technological activity beyond Earth hinges on the possibility that long-lived technosignatures are possible, or even probable. We look more specifically into this issue in the next sections.

\section{Classifying longevity}\label{4}

Having established the importance of longevity in the search for technosignatures, we can explore how many technosignatures should exist, depending on their longevity, in order to have at least one detectable in our search volume. 

To fix the ideas and guide the discussion, we can classify technosignatures into a broad scheme based on longevity:
\begin{itemize}
\item Type A: technosignatures that last for a duration comparable to the typical timescale of technological and cultural evolution on Earth, $\tau \sim 10^3$ years 
\item Type B: technosignatures that last for for a duration comparable to the typical timescale of biological evolution of species on Earth, $\tau \sim 10^6$ years 
\item Type C: technosignatures that last for for a duration comparable to the typical timescale of stellar and planetary evolution, $\tau \sim 10^9$ years 
\end{itemize}

Unlike the famous Kardashev's energy-based classification \citep{Kardashev1964}, this is not meant to classify civilizations, but only their technosignatures. Thus, in this context, longevity should be intended in a general sense, and it may or may not coincide with the survival time of a technological species: rather, it indicates the persistence of its artifacts or of detectable evidences of its existence. This is a major departure from the classical approach to SETI, where the factor $L$ was explicitly linked to the duration of communicating civilizations. Also, it extends the scope of the search to locations that are not necessarily identified with planetary systems: for example, an interstellar probe emitting a detectable technosignature would be classifiable according to the same scheme, and could be treated along the lines discussed in the previous section. One should note that the distribution in Figure~\ref{age} demonstrates an analogy with crossing times for the Galaxy in scenarios with colonizing civilizations or self-replicating probes \citep[e.g.,][]{Sagan1983, Barlow2013}. These considerations have been invoked in discussion of Fermi's paradox, and have been considered disturbing since the median times for colonization of about $10^8$ yrs are still uncomfortably smaller than the age of the Galaxy \citep{Cirkovic2018}. 

Another appealing feature of the proposed classification is that it allows one to treat technosignatures as generic astronomical sources, partially decoupling their phenomenology from the complex bio-sociological trajectories of the civilizations that created them. For example, while some long duration technosignatures can only be produced by civilizations that are high in the Kardashev scale (notably, those requiring energy-intensive planetary engineering), others can easily be left over by societies in the early stages of their technological development (for example, persistent atmospheric pollutants, space debris, or chemically propelled interstellar probes). Even a relatively short-lived civilization low on Kardashev scale can, in principle, produce a large number of technosignatures; launching many cheap interstellar probes is one manner of doing so. It is likely that development of even modest Solar System industrial infrastructure in the course of the next century or so will demonstrate many additional ways of doing so which are unconceived at present. Also, the same civilization can produce technosignatures of various types, over the course of its history. For instance, our species has not yet produced Type A technosignatures, if we only consider the leakage of radio transmissions or the alteration of atmospheric composition by industrial activity; but its artifacts, such as the Voyager 1 and 2, Pioneer 10 and 11, and New Horizons probes, could in principle become type B or even C in the far future, even if our civilization should not survive that long. Similarly, a Type C technosignature can equally be produced by a very long-lived civilization, or by one that has gone extinct on a shorter time scale but has left behind persistent remnants, such as a beacon in a stable orbit or a Dyson-like megastructure. 

\begin{table*}
\caption{Minimum total number $N_{\rm tot}$ of technosignatures of different longevity (classified by Type: see text) and different temporal distribution (Model 1, 2 and 3: see text) that had to appear over galactic history ($T_G=10^{10}$ Gy) in a sphere of radius $R_E=10^3$ ly around Earth, in order to have an expected number $N=1$ detectable today.}\label{ntot}
\label{table}
\begin{center}
\begin{tabular}{llll}
\tableline
 \ & Type A &Type B & Type C  \\
\ & ($\tau=10^3$ y) & ($\tau=10^6$ y) & ($\tau=10^9$ y)\\
\tableline

Model 1: uniform, $0\le t_i \le T_G$ & $N_{\rm tot}=10^7$ & $N_{\rm tot}=10^{4}$ & $N_{\rm tot}=10$ \\ 
Model 2: normal, $\overline{t_i} $=1.5 Gy, $\sigma_{t_i}$=2 Gy  & $N_{\rm tot}=6 \times 10^6$ & $N_{\rm tot}=5 \times 10^{3}$ & $N_{\rm tot}=5$ \\ 
Model 3: normal, $\overline{t_i} $=0 Gy, $\sigma_{t_i}$=2 Gy  & $N_{\rm tot}=3 \times 10^6$ & $N_{\rm tot}=3 \times 10^{3}$ & $N_{\rm tot}=3$ \\ 
\tableline
\end{tabular}
\end{center}
\end{table*}

Getting back to the question of how abundant technosignatures should be in order to be detectable, Table~\ref{ntot} shows the minimum total number of technosignatures of each type that should have appeared, over the course of galactic history, in order to have at least one of them in causal contact with us at present. In the classic Drake equation, this amounts to setting $N=1$ for a given average longevity $L$, so that $N_{\rm tot}=T_G/L$. This corresponds to the result of our simulations for Model 1, i.e.\ for a uniform distribution in $t_i$ (first row of Table~\ref{ntot}). However, two important departures of our simulations from the standard treatment must be highlighted. First, assuming that $t_i$ is not uniformly distributed changes the outcome. As intuitively expected, if there is preference for values of $t_i$ closer to the present epoch, then a smaller $N_{\rm tot}$ is needed in order to have at least one detectable technosignature. Still, $N_{\rm tot}$ needs to be large for short-lived technosignatures, regardless of their distribution. In particular, for Type A technosignatures, $N_{\rm tot}$ has to be of order $\sim 10^7$, comparable to the total number of stellar systems within $R_E$, meaning that {\em every star} should harbor technosignatures if we have to detect one. This is clearly unrealistic, and reiterates our main point: short-lived technosignatures are extremely unlikely to be detectable. The situation improves for Type B, where only $\sim 10^3$ technosignatures are needed, and it gets even better for Type C: only a handful of them, distributed over galactic history, would be enough to have one in contact with us. 

This introduces the second important departure from the classic SETI treatment. Because the factor $L$ in the Drake equation is an {\em average} quantity, one can easily be caught into a defeatist attitude, because large average longevities may arguably seem, a priori, more unlikely than small ones. However, this obscures the fact that we don't really need a {\em large number} of long-lived technosignatures to detect one. For example, if the actual distribution of $\tau$ is dominated by a large number of Type A technosignatures, with only very few Type C outliers, our results show that we would have good chances of detecting one of those long-lived ones.   There is some irony in this conclusion, since for most of the last 60+ years our SETI efforts have been strongly influenced by the hypotheses and models implying a large value of Drake's $N$. Those have been considered ``optimistic'', from the point of view of practical success of SETI. 

Furthermore, it is important to stress that the existence of a few long-lived technosignatures does not presuppose the necessity of an underlying large population of long-lived exo-civilizations, nor it is directly related to the longevity of the civilizations themselves. As noted elsewhere \citep{Cirkovic2018}, the emphasis of classical SETI on large values of $N$ is a red herring. In order for our search to succeed, it is enough that a few civilizations (perhaps even {\em just one}) appearing over the history of the Galaxy managed to leave behind long-duration (i.e.\ Type C) remnants of their presence. One way of actualizing such a possibility is to envision a scenario of directed panspermia: an early, long-living civilization seeded many potentially habitable planets all over the Galaxy \citep{Crick1973}. In such a view, other biospheres would essentially be identical with the technosignatures of the original ``seeder'' civilization.   Directed panspermia is generally considered to be a far-fetched idea at best, as far as abiogenesis on Earth is concerned. Again, if we abandon the strictures of temporal Copernicanism \citep{Cirkovic2020}, this form of panspermia may become more interesting, since its rate is likely to increase with cosmic time, even if it was improbably low in the past.

More generally, long-lived technosignatures can be the natural outcome of any civilization that survives beyond a minimal threshold. In fact, in the next section we propose some theoretical ideas on how this scenario could be realized in practice. 

\section{Modeling longevity}\label{5}

Until now, we did not say anything about what might be a plausible probability distribution of $\tau$. To do that, we would need a {\em theory} of exo-civilizations, which is far above what we can actually afford. However, we can still explore the issue phenomenologically, drawing inspiration from what we learned from the study of the longevity of complex systems on Earth. 

First, we notice that the required number of technosignatures of the various longevity types  that should exist in order to have at least one detectable today follows a power law $\sim \tau ^{-\alpha}$ with $\alpha = 1$  (see Table~\ref{ntot}). How plausible is this? Power-law distributions are very common in nature, and appear in the description of a wide range of phenomena involving ranking, such as city population, number of citations, earthquakes magnitude, wealth distribution and so on \citep[see, e.g.,][]{Newman2005}. Admittedly, however, a value $\alpha =1$ is quite extreme, as for $\alpha\le 2$ the mean of the distribution formally diverges (the famous 80/20 Pareto rule originates from $\alpha\simeq 2.1$).

However, there is no strict need for such a distribution of longevities, as Type B or C technosignatures could be produced in other ways. The simplest is to assume an abrupt transition in longevity: once an exo-civilization reaches a certain technological level, it can create technosignatures that will last for $\tau \gtrsim 10^6$, even if the civilization itself is much younger than that (and even if it goes extinct after some time). For example, one can suppose that a civilization can start creating long-lasting technosignatures after a few thousand years from the moment when it appears (for example, when it reaches Kardashev 2 level). In this scenario, an initial population of civilizations with an average $\tau$ of order $\sim 10^3$ can produce a number of long-lived technosignatures large enough to be detectable. 

This phase-transition scenario can be made more detailed by introducing a time varying termination rate---that is, by relating the additional life expectancy of a technosignature to its age. A function that proved very useful to this extent is the Weibull probability distribution \citep{Weibull1951,Papoulis2002}:
\begin{equation}
p(t;k,t_s)=\frac{k}{t_s}\left(\frac{t}{t_s} \right)^{k-1}e^{-(t/t_s)^k}
\end{equation}
with $k>0$ and $t_s>0$. This provides a flexible mathematical framework to describe a variety of situations in survival analysis, from living organisms to artificial systems. It was also used to model the longevity of technological innovations, cultural products, and so on. The random variable $t>0$ represents the time before termination. The rate of termination, also called `hazard function', is controlled by the shape parameter $k$. Specifically, the termination rate is a power of time, 
\begin{equation}
h(t)= \frac{k}{t_s}\left(\frac{t}{t_s}\right)^{k-1}
\end{equation}
so that \citep{Jiang2011}:
\begin{itemize}
\item If $k >1$, the termination rate increases with time. This is useful for parameterizing aging (for example in living systems), since there will be an increasing probability of termination as time passes (the old is more probable to terminate than the young);
\item If $k=1$, the termination rate is constant, and the Weibull distribution becomes an exponential distribution with average equal to $t_s$. This describes situations where there is no relation between the age of a system and its life expectancy, as when termination is caused by random external events (old and young die with the same probability);
\item If $k <1$, the termination rate decreases with time, resulting in a fat-tailed distribution, similar to Pareto's power-law. This is a common behaviour observed in the diffusion of technology or ideas: the mortality of innovations is initially high but, once a technology has been widely adopted, and has survived for some time, it is likely to survive for at least as much---a phenomenon known as `Lindy effect' (the young is more probable to terminate than the old \citep{Mandelbrot1984,Taleb2012}. 
\end{itemize}

\begin{figure}[ht]
\includegraphics[width=\columnwidth]{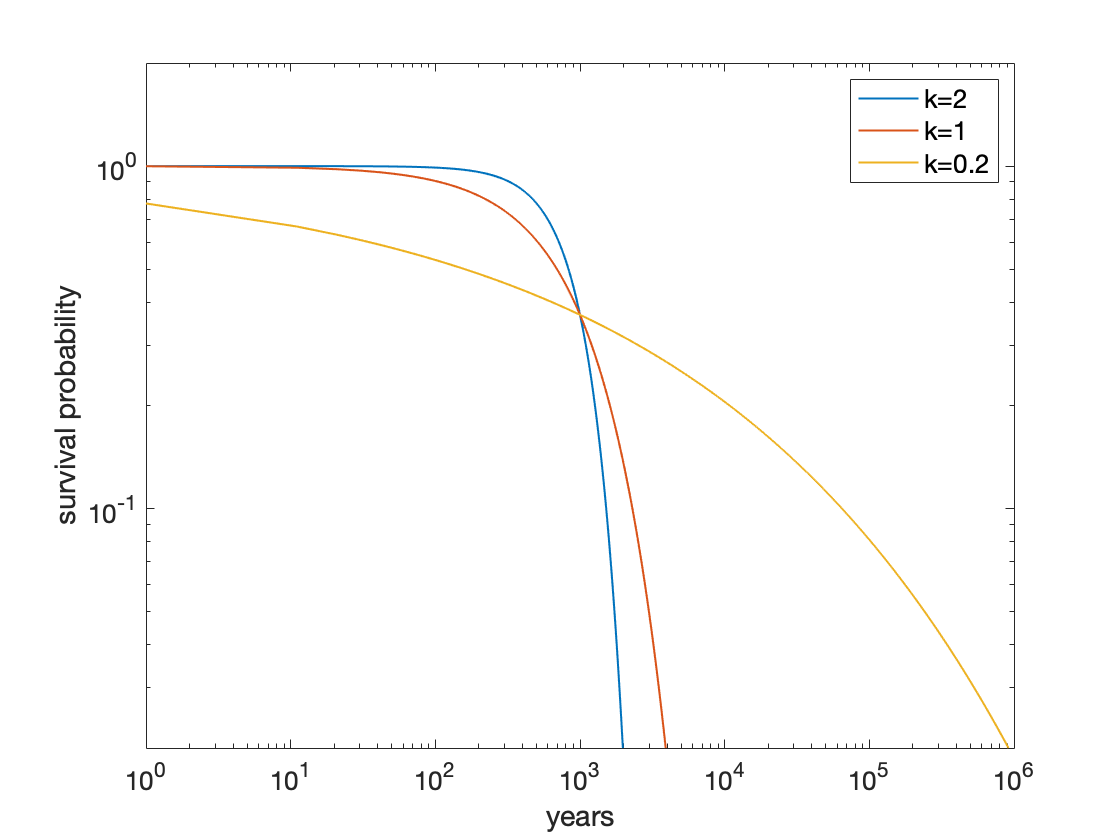}
		\caption{Survival function for a Weibull distribution with $t_s=10^3$ years, for three values of $k$ corresponding to increasing ($k>1$), constant ($k=1$) and decreasing ($k<1$) termination rate.}\label{survival}
\end{figure}

The {\em survival function}, that is the probability that the longevity will be above a certain value $\tau$, is:
\begin{equation}
S(\tau)=\int_{\tau}^\infty  p(t)dt = e^{-(\tau/t_s)^k}
\end{equation}
This is shown in Fig.~\ref{survival} for different values of $k$.  

\begin{figure}[ht]
\includegraphics[width=\columnwidth]{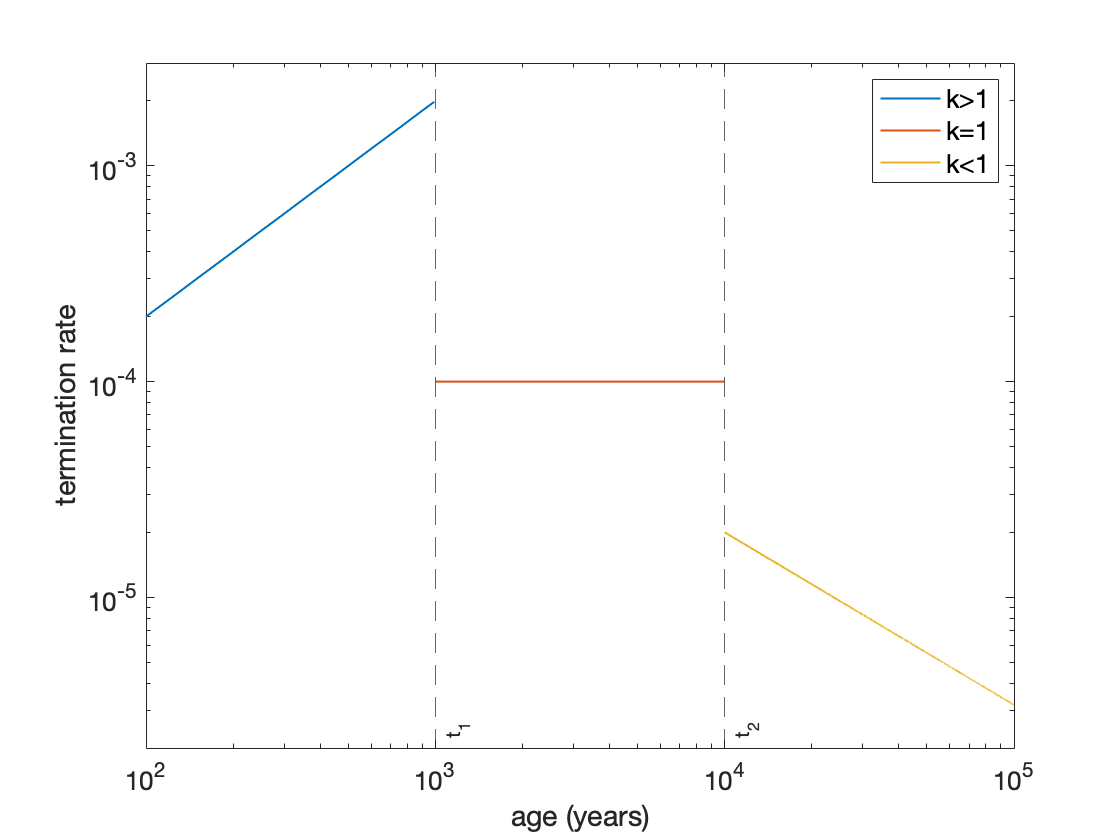}
		\caption{Termination rate for a double transition model in which the probability that civilizations ends increases as their age approaches $t_1=10^3$ years, remains constant for ages between $t_1$ and $t_2=10^4$ years, and decreases for larger ages.}\label{termination}
\end{figure}

This framework allows one to model various scenarios for the longevity of technosignature. A potentially interesting one might involve two epochs of transition (Fig.~\ref{termination}). First, for ages smaller than some $t_1$, civilizations have high mortality rates, because they face a number of obstacles or existential threats: the termination rate increases as time passes ($k<1$). Then, if they manage to survive after a time $t_1$, they may enter a phase where the termination rate is roughly constant ($k=0$), and only depends on external events (e.g.\ some random unavoidable global catastrophe). This may eventually lead to another transition at time $t_2$, after which the termination rate decreases with time, with longevity increasing drastically, as would be expected in the passage from organic to artificial `life' (often denoted as ``postbiological transition''; e.g., \citet{Moravec1999}), or if persistent artifacts can be produced.  Other examples of sharp longevity transitions would include the ability to engineer long-lived, error-free autonomous probes or to maintain terraformed worlds in steady states for eons. There is an often downplayed social element here, since most long-lived technosignatures and advanced engineering feats in general are likely to require high levels of social stability and robustness. Achieving such form of institutions and social structures might count as an advanced engineering feat in its own right. Once again, if longevity is intended as the duration of technosignatures, it is not unplausible that the termination rate decreases with age, as this is commonly observed behavior for technologies that reach a certain degree of maturity.

\begin{figure}[ht]
\includegraphics[width=\columnwidth]{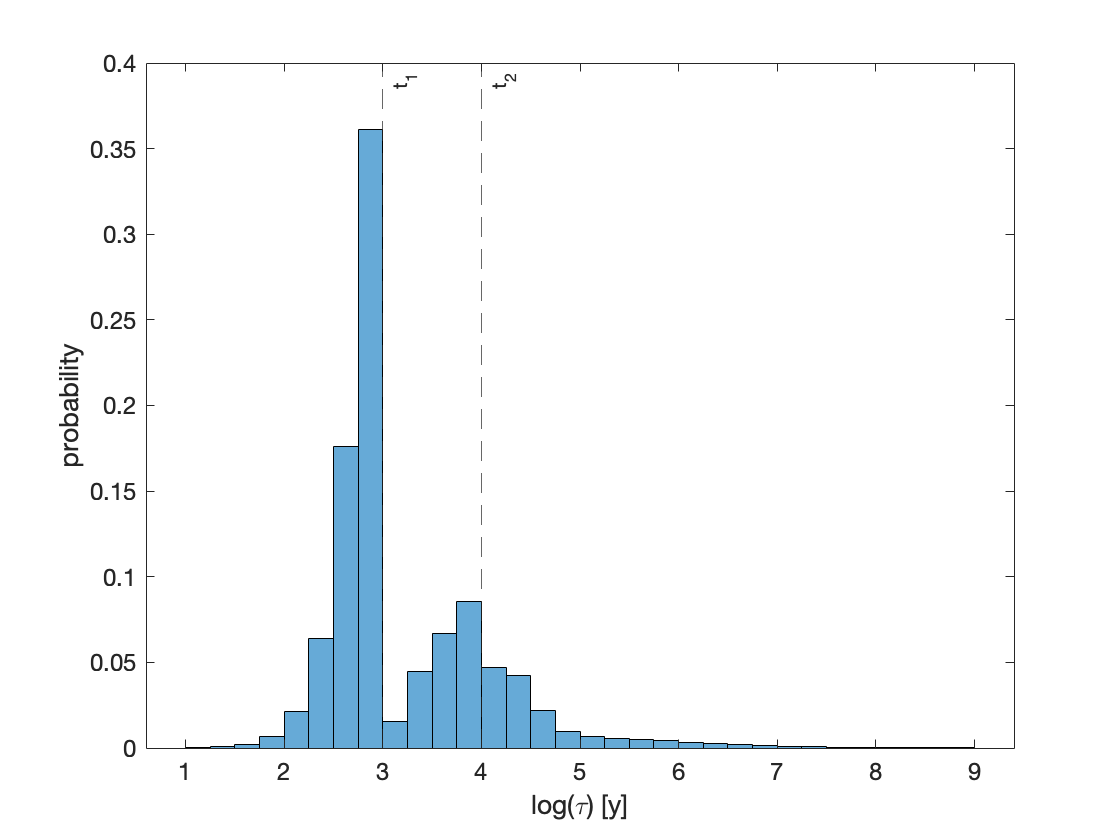}
		\caption{The distribution of longevities $\tau$ resulting from the double transition model of Eq.~\ref{double-transition}.}\label{transition}
\end{figure}

Fig.~\ref{transition} shows $10^7$ longevities simulated according to this double-transition model, with $t_1=10^3$ years and $t_2=10^4$ years, and Weibull distributions having 
\begin{equation}
(k,\ t_s)=\left\{ 
\begin{array}{ll}
(2,\ t_1), & \mbox{if $t\le t_1$} \\
(1,\ t_2), & \mbox{if $t_1<t\le t_2$} \\
(0.2,\ t_2), & \mbox{if $t > t_2$}
\end{array}\label{double-transition}
\right.
\end{equation}
The resulting distribution for $\tau$ is such that, although most longevities fall around Type A (the median of the distribution is $\sim 800$ years) the tail of the distribution is very long, so that there is a large number of Type B and C technosignatures (more than $\sim 10^5$ have $\tau >10^6$ years). This is enough to significantly increase the chances of having detectable technosignatures. When we simulate technosignatures at random locations within $R_E=10^3$ light years, with $t_i$ distributed according to the previously discussed Model 2, and $\tau$ distributed according to the double-transition model, the minimum total number of technosignatures needed to have at least one in causal contact with us is $N_{\rm tot}\sim 3\times 10^4$. This implies that if  $\sim 1/1000$ stellar systems in the surveyed volume have harbored technological species at some point in the history of the Galaxy, we might find proof of their existence. Given that the empirical fraction of potentially habitable planets in the galaxy is $\sim 0.2$ \citep{Petigura2013}, this may not be an implausible expectation.

We emphasize that the model presented in this section is not meant to be a realistic prediction of the actual evolutionary trajectory of technological civilizations, nor it exhausts the space of possibilities. Rather, it serves as an illustration of the non-obvious consequences that can result from a more careful examination of longevity. 

\section{Conclusions}

We examined the longevity of technosignatures and how this factor affects the probability of success in the search for extraterrestrial intelligence. In early SETI, longevity had a very restricted meaning and essentially was synonymous with the overall lifetime of technological civilizations. Therefore, although the importance of the factor $L$ in the Drake equation was recognized from the very beginning, the subtleties involved in its definition and in the underlying statistical features were not explored in detail. This resulted in a situation in which an empirical success of SETI observations seemed predicated on a large number of coexistent Galactic civilizations at a vaguely similar stage of development (the famous ``Galactic Club''). The improbability of such state of affairs has been masterfully exploited by opponents of SETI, resulting not only in well-known funding issues, but also in widespread confusion surrounding the whole enterprise. In this paper, we followed a more general approach, and looked at longevity under a wider angle, decoupling the lifetime of technosignatures from that of the species that produced it. We also proposed a broad classification scheme for technosignatures, based on their duration, that can be useful to guide discussions and strategies for future searches. 

A major conclusion of our work is that any Galactic technosignature that we can detect at present is most likely very long-lived, with a duration $\gtrsim 10^6$ years. This is irrespective of the underlying distribution and abundance of technosignatures in the Galaxy. Short-lived ($\sim 10^3$ years) technosignatures can be detectable only if they are extremely common (at least one per each stellar system in our neighborhood) or if some global astrophysical mechanism synchronized their appearance at late times in the history of the Galaxy, so that they are essentially coeval with us. This suggests that an anthropocentric approach to SETI is flawed: it is rational to expect that the kind of technosignatures we are most likely to get in contact with is wildly different, in terms of duration, from what has been produced over the course of human history. This conclusion strengthens the case for the hitherto downplayed hypothesis (which is not easily labeled as ``optimistic'' or ``pessimistic'') that a significant fraction of detectable technosignatures in the Galaxy are products of extraterrestrial civilizations which are now extinct.

Our work suggests that the most effective search strategy should be focused on extremely long-lived technosignatures (Type B or C in our classification scheme). While the question of how to optimize SETI observations is complicated by the number of parameters involved (for a recent proposal of the relevant degrees of freedom to consider, see \citet{Sheikh2019}) we argue that technosignature longevity is paramount and should have top priority over other factors. Thus, Dysonesque megastructures, interstellar probes, persistent beacons---as well as activities related to civilizations above Type 2 of the Kardashev scale, or to artificial intelligence---should be the preferred target for future searches. These technosignatures would not only be `weird' when measured against our own bias, but could arguably be less common than short-lived ones. Such conclusion deflates the emphasis on large $N$ (and human-like technosignatures) that informed much of classical SETI's literature. However, the supposed rarity of long-lived technosignatures should not be regarded, in itself, as a hindrance for the SETI enterprise: in fact, a few Type C technosignatures, over the course of the entire history of the Galaxy, would have much higher chance of being detected than a large number of Type A.  Also, possible astrophysical mechanisms which could lead to a posteriori synchronization of shorter lived technosignatures should be investigated, to constrain the parameter space of this possibility, if nothing else.

Whether there are evolutionary pathways that lead to the existence of ETIs capable of producing long-lived evidences is an open problem that would deserve closer attention. In principle, technosignature ``interstellar archeology'' could look for relatively primitive and short-lived civilizations that managed---voluntarily or not---to leave persistent testimony of their existence before their demise. At the same time, civilizations that appeared much earlier than ours and that managed to solve their existential crisis may have reached the capacity to last almost indefinitely, and therefore be still active at present. In our work, we gave a first sketch of a possible model of long duration technosignatures based on the transition from an initial high-mortality rate to a more stable long-term behaviour. This should be seen as a prompt for further explorations of more complex and detailed models, whose outcome should not only orient the direction of technosignature studies, but would also be relevant for the future prospects of humanity. 

\acknowledgments Very useful comments and criticisms of the reviewer have resulted in significant improvement of a previous version of this manuscript. A.B. acknowledges useful discussions with Manasvi Lingam and Claudio Grimaldi, as well as partial support by grant number FQXi-MGA-1801 and FQXi-MGB-1924 from the Foundational Questions Institute and Fetzer Franklin Fund, a donor advised fund of Silicon Valley Community Foundation.  M.M.Ć. wishes to thank Anders Sandberg and Branislav Vukotić on many relevant debates (and many disagreements!) on the topic of this paper; he also acknowledges partial support of the Ministry of Education, Science and Technological Development of Serbia under the contract 451-03-9/2021-14/200002.

\end{document}